\documentclass[fleqn,12pt]{wlscirep}
\usepackage{amsmath}
\usepackage{graphicx}
\usepackage{footmisc}
\usepackage{color}

\begin{document}
\title {Fractional quantum Hall effect in the Hofstadter model of interacting fermions}
\author[1]{Igor N. Karnaukhov}
\affil[1]{G.V. Kurdyumov Institute for Metal Physics, 36 Vernadsky Boulevard, 03142 Kiev, Ukraine}
\affil[*]{karnaui@yahoo.com}
\begin{abstract}
Applying a unified approach, we study integer quantum Hall effect (IQHE) and  fractional quantum Hall effect (FQHE) in the Hofstadter model with short range interaction between fermions. An effective field, that takes into account the interaction, is determined by both the amplitude and phase. Its amplitude is proportional to the interaction strength, the phase corresponds to the minimum energy. In fact the problem is reduced to the Harper equation with two different scales: the first is a magnetic scale (cell size corresponding to a unit quantum magnetic flux), the second scale (determines the inhomogeneity of the effective field) forms the steady fine structure of the Hofstadter spectrum and leads to the realization of  fractional quantum Hall states.
In a sample of finite sizes with open boundary conditions, the fine structure of the Hofstadter spectrum also includes the fine structure of the edge chiral modes. The subbands in a fine structure of the Hofstadter band (HB) are separated extremely small quasigaps.
The Chern number of a topological HB is conserved during the formation of its fine structure.  Edge modes are formed into HB, they connect the nearest-neighbor subbands and determine the fractional conductance for the fractional filling at the Fermi energies  corresponding to these quasigaps.
\end{abstract}
\maketitle

\section*{Introduction}

The Harper-Hofstadter model \cite{Har, Hof} plays a key role in the modern understanding and description of topological states on a 2D lattice. It allows us to describe the nontrivial behavior of fermions in an external magnetic field with their arbitrary density and dispersion, determine the structure of topological bands, calculate the Chern numbers in a wide range of magnetic fluxes. For rational magnetic fluxes penetrating a magnetic cell with size $q$ ($q$ is defined in units of the lattice spacing), the Hofstadter model has exact solution \cite{w,w1}. In experimental realizable magnetic fields, which corresponds to semi-classical limit with a magnetic scale $q\simeq 10^3- 10^4$, the spectrum of quasiparticle excitations  is well described in the framework of the Landau levels near the edge spectrum \cite{K1,K2,K3}, and the Dirac levels in graphene \cite{a3,a4,K4}. Irrational magnetic fluxes can be experimental realized only in the samples of small sizes when the size of a sample N is less than a magnetic scale q \cite{K3}. In this case, the q value  is  the maximal scale in the model.

IQHE is explained in the framework of the Hofstadter model \cite{B1,B2,
K1,K2,K3,a3,K4}, the same cannot be said about FQHE. Unfortunately the model is unable to explain FQHE, because it does not take into account the interaction between quantum particles.
FQHE is not sensitive to spin degrees of freedom, so the repulsion between fermions should be taken into account first.
A theory that could explain all the diversity of the FQHE is still lacking, and the nature of the FQHE remains an open question in condensed matter physics. Let's pay tribute to the ideas \cite{A1,A2}, that make it clear that the effect itself is non-trivial.

The purpose of this work is not to explain the numerous experimental data on the measurement
fractional Hall conductivity, we are talking about understanding the physical nature of the FQHE, explain, it would seem,
the controversial concept of fractional conductance.

\section*{Model Hamiltonian and method}

We study FQHE in the framework of the Hofstadter model defined for interacting electrons on square lattice with the Hamiltonian ${\cal H}={\cal H}_0  +{\cal H}_{int}$
\begin{eqnarray}
&&{\cal H}_0= -\sum_{\sigma=\uparrow,\downarrow}\sum_{n,m} [a^\dagger_{n,m;\sigma} a_{n+1,m;\sigma} + \exp(2i \pi n\phi) a^\dagger_{n,m;\sigma}  a_{n,m+1;\sigma} + H.c.] -\\ \nonumber
&& \mu\sum_{\sigma=\uparrow,\downarrow}\sum_{j} n_{j;\sigma}-H \sum_{j} (n_{j;\uparrow}-n_{j;\downarrow}),\\
&&{\cal H}_{int} =U \sum_{j} n_{j;\uparrow} n_{j;\downarrow},
 \label{eq-H1}
\end{eqnarray}
where $a^\dagger_{n,m;\sigma} $ and $a_{n,m;\sigma}$ are the fermion operators located at a site $j=\{n,m\}$ with spin $\sigma =\uparrow,\downarrow$, $n_{j;\sigma}=a^\dagger_{j;\sigma}a_{j;\sigma}$ denotes the density operator,
$\mu$ is a chemical potential. The Hamiltonian ${\cal H}_0$ describes the hoppings of fermions between the nearest-neighbor lattice sites. A magnetic flux through the unit cell $\phi = \frac{H }{ \Phi_0}$ is determined in the quantum flux unit ${\Phi_0=h/e}$, here $H$ is a magnetic field and  a lattice constant is equal to unit. ${\cal H}_{int}$ term is determined by the on-site Hubbard  interaction $U$.

The interaction term (2) can be conveniently redefined in the momentum representation
${\cal H}_{int} = V U  \sum_{\textbf{K}}n_{\textbf{K};\uparrow} n_{-\textbf{K};\downarrow}$,
where $n_{\textbf{K};\sigma}=\frac{1}{V}\sum_j exp(i\textbf{K j})n_{j;\sigma}$, the volume is equal to $V=N \times N$. Using the mean field approach, we rewrite this term as follows ${\cal H}_{int} =V(\lambda_{\textbf{K};\uparrow}n_{\textbf{-K};\downarrow}+\lambda_{\textbf{-K};\downarrow}n_{\textbf{K};\uparrow})$ with an effective field $\lambda_{\textbf{K};\sigma}=U <n_{\textbf{K};\sigma}>$, which is determined by a fixed value of the wave vector $\textbf{K}$.
In the experiments the magnetic fields  correspond to the semi-classical limit with a magnetic scale $q \sim 10^3-10^4$, which corresponds to small values $K\sim 10^{-3}-10^{-4}$. The density of fermions for the states near the low energy edge of  the spectrum is small $\sim \frac{1}{q}$.
In the small K-limit the expression for $\lambda_{\textbf{K};\sigma}$ is simplified  $\lambda_{\textbf{K};\sigma}=\lambda_\sigma +0(K^2)$, where $\lambda_\sigma =U\rho_\sigma $, $\rho_\sigma$ is the density of electrons with spin $\sigma$.
The Zeeman energy shifts the energies of electron bands with different spins, removes spin degeneracy, and does not change the topological state of the electron liquid. This makes it possible to explicitly disregard the dependence of the electron energy on the spin and to consider the problem for spinless fermions. The model is reduced to spinless fermions with the interaction term  ${\cal H}_{int} =\frac{\lambda}{2}\sum_{j}[\exp(i\textbf{K j})+\exp(-i\textbf{K j)}]n_j=\lambda \sum_{j}\cos(\textbf{K j})n_j$, with $\lambda =U\rho$, $n_j$ and  $\rho$ are the density operator of spinless fermions and their filling.

We study the 2D system in a hollow cylindrical geometry with open boundary conditions (a cylinder axis along the $x$-direction and the boundaries along the y-direction). The Hamiltonian ${\cal H}_0 $ describes chains of spinless fermions oriented along the $y$-axis ($n$ is a coordinate of the chain in the x-direction) connected by single-particle tunneling with tunneling constant equal to unit. The wave function of free fermions in the y-chains, which is determined by wave vector $k_y$ are localized in the x-direction \cite{1,3}(for each $k_y$). The amplitudes of the wave function with different values $k_y$ overlap in the x-direction, the eigenstates of the Hamiltonian ${\cal H}_0 $  are the Bloch form. All states with different $n$ are bounded via a magnetic flux. At a low fermion density, the repulsion between quasi-free fermions in y-chains may be ignored, while the repulsion between fermions localized in the x-direction dominates. Thus we can assume that the vector $\textbf{K}$ has only one projection along x $K_x$, note as $K$ or $\textbf{K}=(K,0)$. Making the Ansatz for the wave function $\psi(n,m)=\exp(i k_y m)g_n$ (which determines the state with the energy $\epsilon$) we obtain the Harper equation for the model Hamiltomnian (1),(2)
\begin{equation}
\epsilon g_n =-g_{n+1}-g_{n-1}-2\cos(k_y +2\pi n \phi)g_n +\lambda \cos(K n)g_n.
 \label{eq-Har}
\end{equation}
This equation is key in studying FQHE.

The problem is reduced to the (1+1)D quantum system, where the states of fermions are determined by two phases: the first is  a magnetic phase $\phi$, the second is the phase $K$, which is connected with interaction. The  $K$ value corresponds to minimum energy of the system, it minimizes the energy of electron liquid upon interaction  (2). At T=0K the model is three parameter one, we shall analyze the phase state of the interacting spinless fermions for arbitrary rational fluxes $\phi=\frac{p}{q}$ ($p$ and $q$ are coprime integers), $U$ and $\rho$.
In the Hofstadter model of noninteracting fermions the states of fermions with different $\phi$ are topologically  similar in the following sense: the Chern number, the Hall conductance are determined by the magnetic flux, filling or the number of filled isolated bands that correspond to this filling, while it does not depend on the structure of the subbands \cite{K1,K2,K3} (their bandwidths, the values of the gaps between them).

The effect of the interaction on the behavior of fermions is reduced to the appearance of an inhomogeneous $\lambda$-field, which is determined by the magnitude $\lambda$ and phase $K$, we shall use the following parametrization  $K=2 \pi \frac{r}{s}$, here $r$ and $s$ are relatively prime integers. Such trivial solutions $K=0$ ($s\to\infty$) and $K=2\pi$ ($r=s=1$) correspond to  maximum energy, according to (3) the energy $\varepsilon$ is shifted to the maximal value $+\lambda$. In the $K \to 0$ (or $s\to\infty$) limit the solution for $K $ corresponds to irrational fluxes, that are realized at $s>N$ \cite{K3}. We consider the steady state of the system for rational fluxes, namely for integer $\frac{s}{q}=\alpha$, when $q \leqslant s$, or integer $\alpha^{-1}$, when $q>s$. The minimum energy corresponds to nontrivial solution for $K$ at a given magnetic flux $\phi$.
The fine structure of the Hofstadter spectrum is realized at $\alpha>1$, when the interaction scale is maximal $s>q$.
In the case $\alpha<1$ the spectrum is renormalized, its topological structure remains the same.  We restrict ourselves to considering low energy HB.

\section*{Splitting of low energy Hofstadter bands, \\
a fine structure of the spectrum in the semi-classical limit }

\begin{figure}[tp]
     \centering{\leavevmode}
\begin{minipage}[h]{.4\linewidth}
\center{
\includegraphics[width=\linewidth]{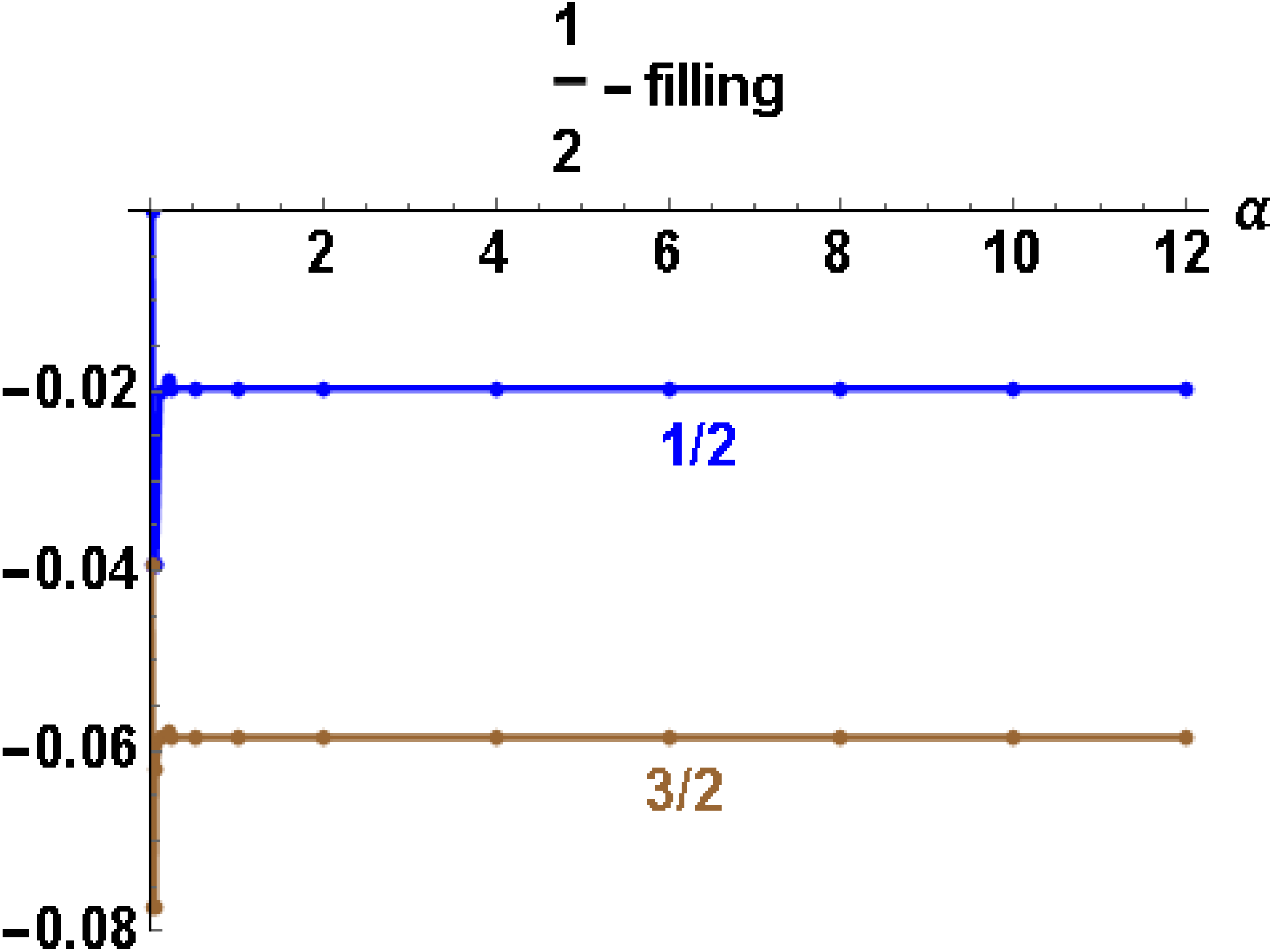} a)\\
                  }
    \end{minipage}
    \begin{minipage}[h]{.4\linewidth}
\center{
\includegraphics[width=\linewidth]{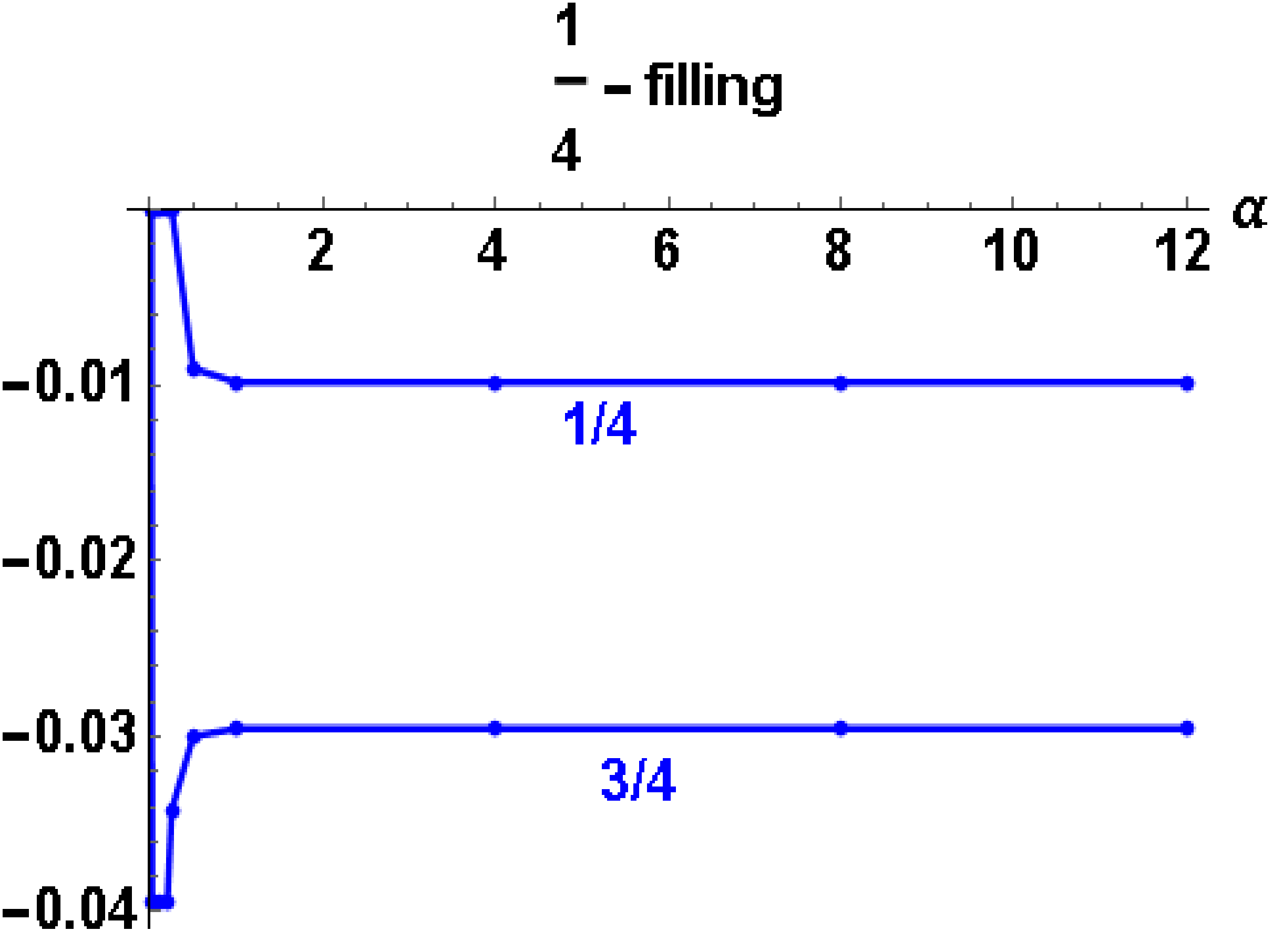} b)\\
                  }
    \end{minipage}
    \begin{minipage}[h]{.4\linewidth}
\center{
\includegraphics[width=\linewidth]{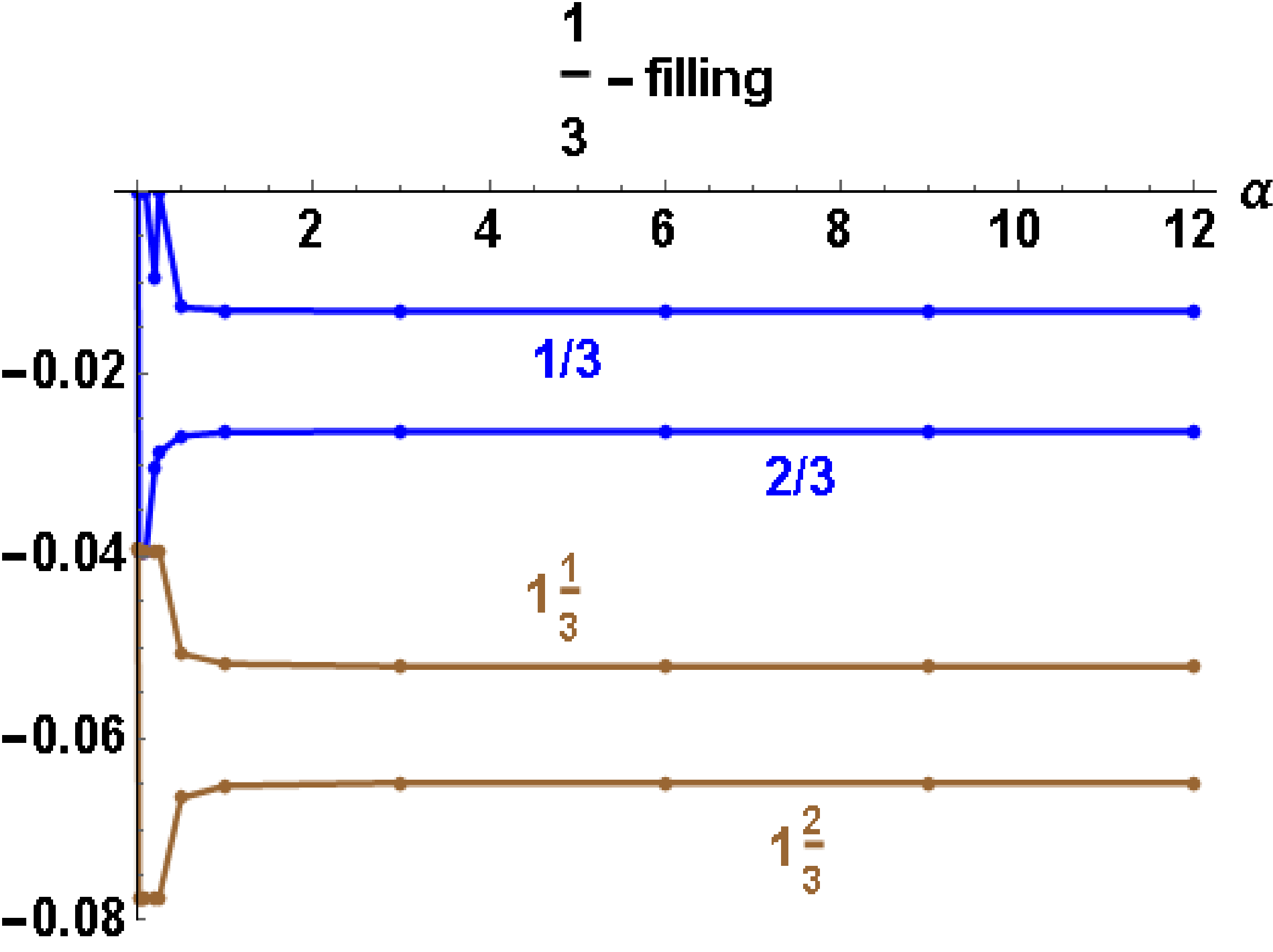} c)\\
                  }
    \end{minipage}
    \begin{minipage}[h]{.4\linewidth}
\center{
\includegraphics[width=\linewidth]{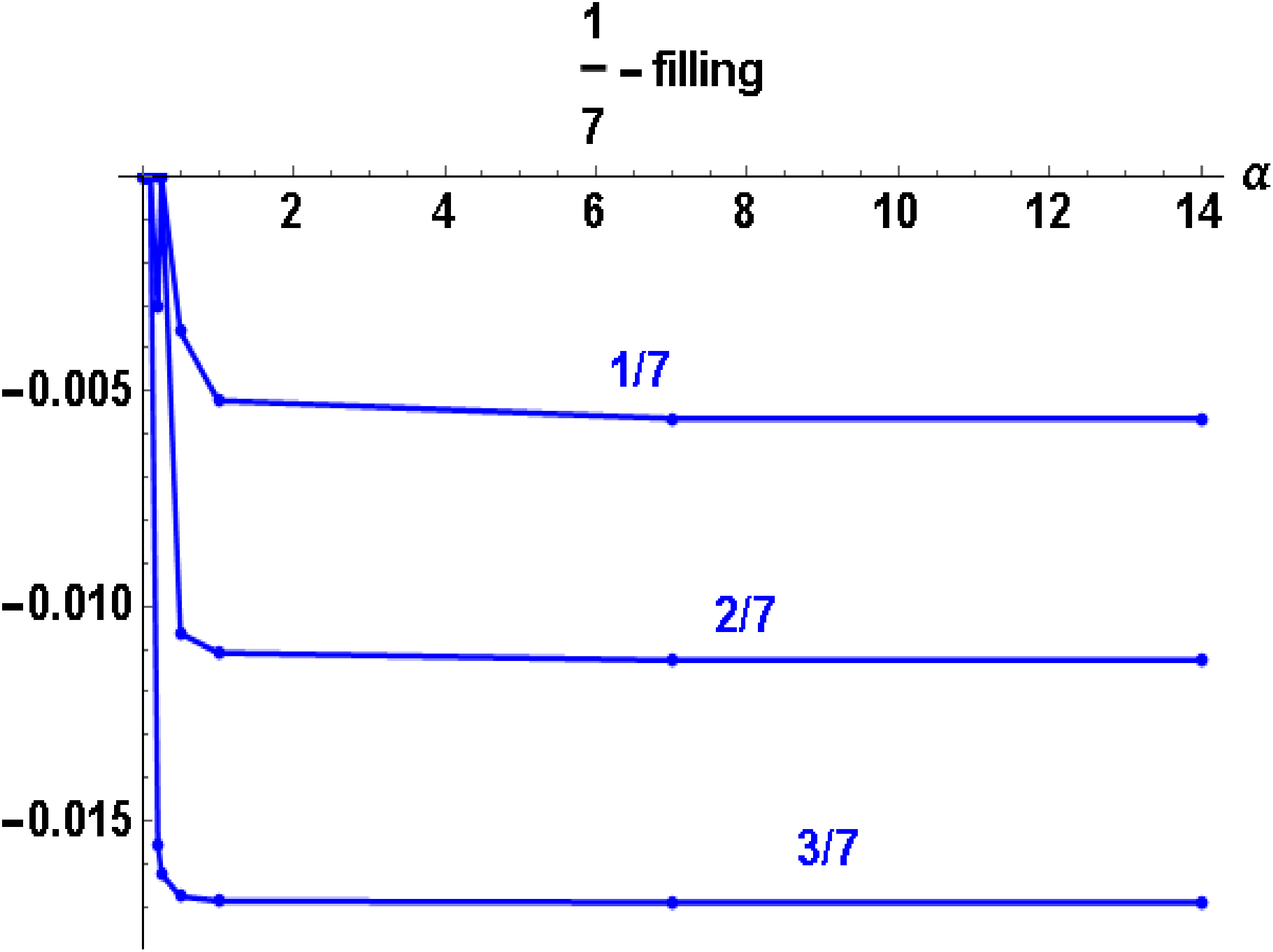} d)\\
                  }
    \end{minipage}
    \begin{minipage}[h]{.4\linewidth}
\center{
\includegraphics[width=\linewidth]{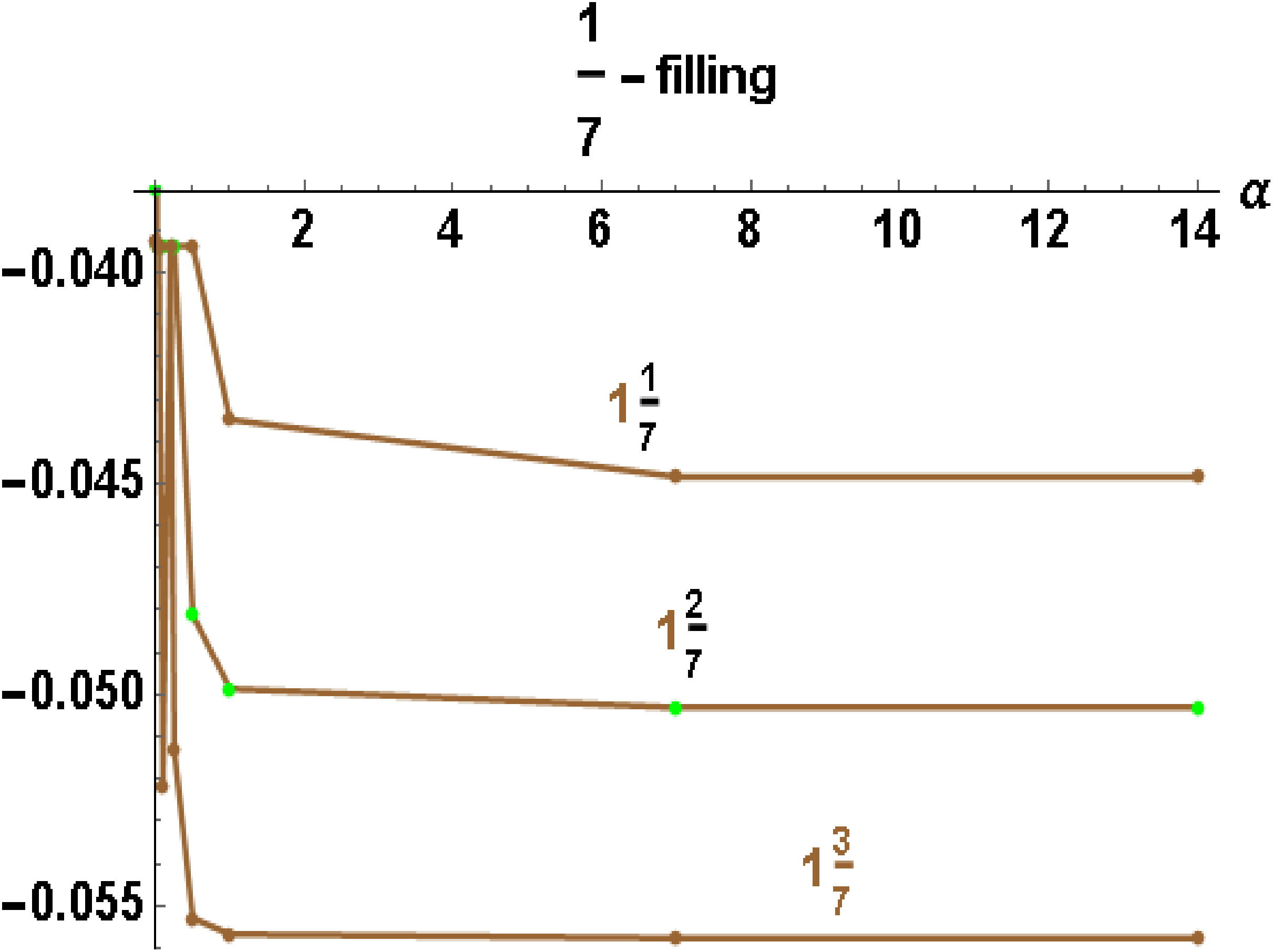} e)\\
                  }
    \end{minipage}
    \begin{minipage}[h]{.4\linewidth}
\center{
\includegraphics[width=\linewidth]{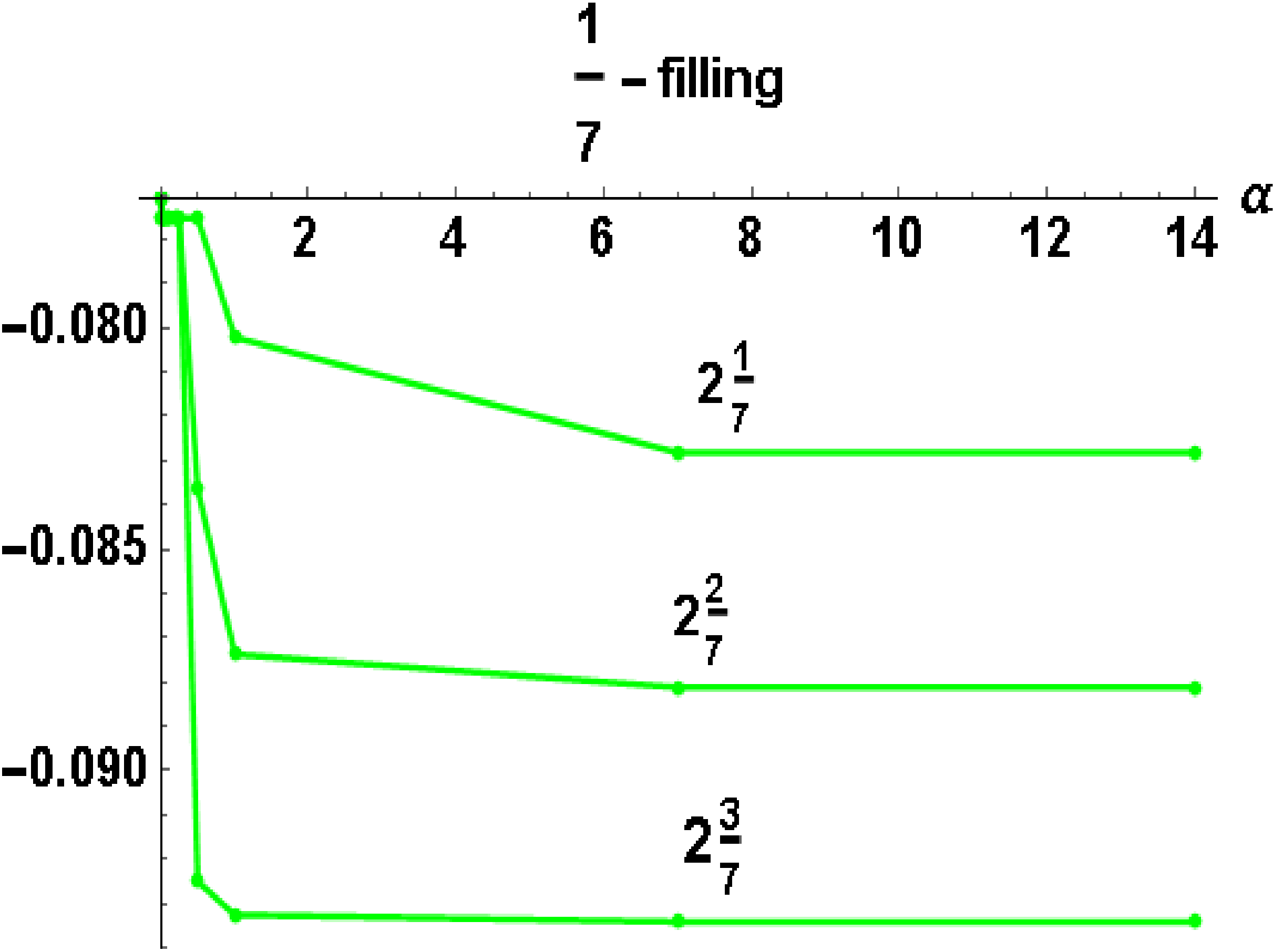} f)\\
                  }
    \end{minipage}
\caption{(Color online)
The energy density as a function of $\alpha$ calculated at $q=10^2, U=1$ for:\\
a) $\frac{1}{2}$-fraction filling for the first
$\nu =\frac{1}{2}$ and second $\nu =1\frac{1}{2}$  HB
(the points $\alpha \geq 2 $ characterize the unstable fine structure of HB);
b) $\frac{1}{4}$-fraction filling for the first HB $\nu=\frac{1}{4}$ and $\nu=\frac{3}{4}$ ($\frac{2}{4}$- or
$\frac{1}{2}$-filling is shown in a)), steady fractional state with $\alpha\geq 4$ and filling $\frac{1}{4}$ is realized also in the second HB;
c)  $\frac{1}{3}$-fraction filling, steady fractional state for $\nu=\frac{1}{3}$ and $\nu=1\frac{1}{3}$, unsteady state  for $\nu=\frac{2}{3}$ and $\nu=1\frac{2}{3}$;
d),e), f)  $\frac{1}{7}$-fraction filling for three HB, the steady states are shown at $\nu =\frac{1}{7}, \frac{2}{7}, \frac{3}{7}$ for the fist HB d),  $\nu =1\frac{1}{7}, 1\frac{2}{7}, 1\frac{3}{7}$ for the second e),  $\nu =2\frac{1}{7}, 2\frac{2}{7}, 2\frac{3}{7}$ for the third f).
}
\label{fig:1}
\end{figure}
We provide numerical analysis of the quasi-particle excitations near the edge of the spectrum considering rational fluxes $\phi$ and $K$   in the semi-classical limit with $p=1$ and $r=1$ for different $q\gg 1$, $s\gg 1$ and filling $\rho \ll 1$. Magnetic fields, at which measurements are carried out, correspond to large $q\sim 10^3-10^4$, so there is no point in considering the case of $q \sim 1$. As a reasonable compromise with numerical calculations (large, but not very large $q$), we consider splitting (due to the interaction) of low energy fermion bands at $q=10^2$. In contrast to the traditional Hofstadler model there are two scales: the first $q$ is determined by an external magnetic field, the second $s$ is associated with the on-site repulsion. For states near the edge spectrum, value of $\lambda$ corresponds to a weak interaction limit, because the filling $\rho \sim \frac{1}{q}$ or $\rho \sim \frac{1}{s}$.

We will show that the fine structure of the spectrum, which is formed due to the on-site repulsion, determines the FQH states in an external magnetic field.  First of all consider the formation of a fine structure of low energy HB, which corresponds to filling less than $ \frac {1} {q} $ for the first (the lowest) HB and when filling $ \frac {1} {q} \leq\rho \leq  \frac {2} {q} $ for the second band, where $ q = 10^2 $ is fixed for numerical calculations.
A rather obvious consequence follows from numerical calculations: in  a weak coupling at $\rho U <1$, that is valid in semi-classical limit for an arbitrary bare value of $U$, the fine structure of the spectrum does not depend on the value of $ \lambda $. This makes it possible to consider the evolution of the fermion spectrum at a fixed value of $U=1$ or $\lambda=\rho$ and different $s$. We fix $ q =10^2$, $U=1$ and calculate the spectrum for various  $ \alpha = \frac{s}{q} $, which corresponds to rational fluxes, when $\alpha$ or $\alpha^{-1}$ is an integer.

It's nice that the spectrum has a fairly simple topologically stable structure. The number of HB in the spectrum is equal to $q$,  at $\alpha>1$  $\alpha$ subbands form a fine structure of each HB.  The values of the gaps between low energy HB $\Delta_{j,j+1}(\alpha)$ ($j$ numerates the band) depend on $q$ and $\lambda$ and insignificantly on $\alpha$, at $q=10^2$ and $U=1$:\\
$\Delta_{1,2}(1)= 0.1043$, $\Delta_{2,3}(1)=0.0842$;\\
$\Delta_{1,2}(2)= 0.1039$, $\Delta_{2,3}(2)=0.0824$;\\
$\Delta_{1,2}(3)= 0.1038$,  $\Delta_{2,3}(3)=0.0820$;\\
$\Delta_{1,2}(4)=0.1037$,  $\Delta_{2,3}(4)=0.0819$;\\
$\Delta_{1,2}(5)=0.1037$, $\Delta_{2,3}(5)=0.0818$; \\
$\Delta_{1,2}(6)=0.1037$, $\Delta_{2,3}(6)=0.0818$; \\
$\Delta_{1,2}(7)=0.1037$, $\Delta_{2,3}(7)=0.0818$. \\
For $\alpha>1$ the values of the gaps in the Hofstadter spectrum are practically independent of the value of $\alpha$,
$\Delta_{1,2}=0.1237$, $\Delta_{2,3}=0.1217$ at $U=0$ in the Hofstadter model, for comparison.

The bandwidths in  the fine structure of the j-HB $\epsilon_{j,i}(\alpha)$ ($i=1,...,\alpha)$ ($j=1,2$ numerates HB and $i$ numerates the subband in j-band) are equal to \\
$\epsilon_{1,1}(1)=0.0197$,  $\epsilon_{2,1}(1)=0.0381$;\\
$\epsilon_{1,1}(2)=0.0050$, $\epsilon_{1,2}(2)= 0.0100$, $\epsilon_{2,1}(2)=0.0148$, $\epsilon_{2,2}(2)= 0.0198$;\\
$\epsilon_{1,1}(3)=0.0017$, $\epsilon_{1,2}(3)=0.0067 $, $\epsilon_{1,3}(3)=0.0050$,
$\epsilon_{2,1}(3)= 0.0066$, $\epsilon_{2,2}(3)= 0.0166$, $\epsilon_{2,3}(3)= 0.0100$; \\
$\epsilon_{1,1}(4)=0.0007$, $\epsilon_{1,2}(4)= 0.0035$, $\epsilon_{1,3}(4)= 0.0053$, $\epsilon_{1,4}(4)=0.0029$,
$\epsilon_{2,1}(4)=0.0036$, $\epsilon_{2,2}(4)=0.0106$, $\epsilon_{2,3}(4)=0.0123$, $\epsilon_{2,4}(4)= 0.0058$; \\
$\epsilon_{1,1}(5)= 0.0004$, $\epsilon_{1,2}(5)= 0.0020$,  $\epsilon_{1,3}(5)=0.0037$, $\epsilon_{1,4}(5)=0.0040$,  $\epsilon_{1,5}(5)=0.0019$,
$\epsilon_{2,1}(5)=0.0023$, $\epsilon_{2,2}(5) =0.0070$,  $\epsilon_{2,3}(5)=0.0099$, $\epsilon_{2,4}(5\sum)=0.0090$,  $\epsilon_{2,5}(5)=0.0038$;\\
$\epsilon_{1,1}(6)=0.0002$, $\epsilon_{1,2}(6)=0.0012$, $\epsilon_{1,3}(6)=0.0025$, $\epsilon_{1,4}(6)= 0.0033$, $\epsilon_{1,5}(6)= 0.0030$, $\epsilon_{1,6}(6)=0.0013$, $\epsilon_{2,1}(6)=0.0016$, $\epsilon_{2,2}(6)=0.0049$,
$\epsilon_{2,3}(6)=0.0075$, $\epsilon_{2,4}(6)= 0.0083$, $\epsilon_{2,5}(6)= 0.0067$, $\epsilon_{2,6}(6)=0.0027$; \\
$\epsilon_{1,1}(7)= 0.0001$, $\epsilon_{1,2}(7)= 0.0008$,  $\epsilon_{1,3}(7)= 0.0017$, $\epsilon_{1,4}(7)= 0.0025$, $\epsilon_{1,5}(7)= 0.0029$,
$\epsilon_{1,6}(7)=0.0024$, $\epsilon_{1,7}(7)= 0.0010$, $\epsilon_{2,1}(7)= 0.0011$, $\epsilon_{2,2}(7)= 0.0036$,  $\epsilon_{2,3}(7)= 0.0057$, $\epsilon_{2,4}(7)= 0.0070$, $\epsilon_{2,5}(7)= 0.0069$, $\epsilon_{2,6}(7)=0.0051$,  $\epsilon_{2,7}(7)= 0.0020$.

The on-site repulsion forms the subbands in the Hofstadter spectrum (note, that in the semi-classical limit the Hofstadter spectrum is reduced to the Landau levels). Narrow subbands with bandwidths $\epsilon(j,i)\ll\Delta(j,j+1)$ form the fine structure of HB, their values increase with increasing the HB number and decrease with increasing $\alpha$.
Quasigaps between subbands i and i+1 in fine structure of the j-HB $\delta\varepsilon_{j;i,i+1}(\alpha)$ are extremal small, so they are \\
$\delta\varepsilon_{1;1,2}(2)\sim 3\centerdot 10^{-11}$, $\delta\varepsilon_{2;1,2}(2)\sim 2\centerdot 10^{-10}$;  \\
$\delta\varepsilon_{1;1,2}(3)\sim 5\centerdot 10^{-12}$, $\delta\varepsilon_{1;2,3}(3)\sim 9\centerdot 10^{-11}$,
$\delta\varepsilon_{2;1,2}(3)\sim 9\centerdot 10^{-13}$, $\delta\varepsilon_{2;2,3}(3)\sim 1\centerdot 10^{-10}$; \\
$\delta\varepsilon_{1;1,2}(4)\sim 4\centerdot 10^{-12}$, $\delta\varepsilon_{1;2,3}(4)\sim 3\centerdot 10^{-11}$,
$\delta\varepsilon_{1;3,4}(4)\sim 6\centerdot 10^{-11}$,
$\delta\varepsilon_{2;1,2}(4)\sim 3\centerdot 10^{-11}$, $\delta\varepsilon_{2;2,3}(4)\sim 9\centerdot 10^{-13}$,  $\delta\varepsilon_{2;3,4}(4)\sim 2\centerdot 10^{-11}$;\\
$\delta\varepsilon_{1;1,2}(5)\sim 1\centerdot 10^{-11}$, $\delta\varepsilon_{1;2,3}(5)\sim 1 \centerdot 10^{-11}$, $\delta\varepsilon_{1;3,4}(5)\sim 2\centerdot 10^{-11}$, $\delta\varepsilon_{1;4,5}(5)\sim 1\centerdot 10^{-11}$,
$\delta\varepsilon_{2;1,2}(5)\sim 4\centerdot 10^{-11}$, $\delta\varepsilon_{2;2,3}(5)\sim 1\centerdot 10^{-10}$, $\delta\varepsilon_{2;3,4}(5)\sim 3\centerdot 10^{-11}$, $\delta\varepsilon_{2;4,5}(5)\sim 2\centerdot 10^{-11}$; \\
$\delta\varepsilon_{1;1,2}(6)\sim 1\centerdot 10^{-11}$, $\delta\varepsilon_{1;2,3}(6) \sim 4\centerdot 10^{-12}$, $\delta\varepsilon_{1;3,4}(6) \sim 3\centerdot 10^{-11}$, $\delta\varepsilon_{1;4,5}(6) \sim 3\centerdot 10^{-11}$, $\delta\varepsilon_{1;5,6}(6) \sim 3\centerdot 10^{-11}$,
$\delta\varepsilon_{2;1,2}(6)\sim 2\centerdot 10^{-11}$, $\delta\varepsilon_{2;2,3}(6) \sim 9\centerdot 10^{-11}$,
$\delta\varepsilon_{2;3,4}(6) \sim 1\centerdot 10^{-10}$, $\delta\varepsilon_{2;4,5}(6) \sim 3\centerdot 10^{-11}$,
$\delta\varepsilon_{2;5,6}(6) \sim 5\centerdot 10^{-11}$;\\
$\delta\varepsilon_{1;1,2}(7)\sim 3\centerdot 10^{-12}$, $\delta\varepsilon_{1;2,3}(7)\sim 3\centerdot 10^{-11}$,
$\delta\varepsilon_{1;3,4}(7)\sim 8\centerdot 10^{-12}$,  $\delta\varepsilon_{1;4,5}(7)\sim 4\centerdot 10^{-11}$,
$\delta\varepsilon_{1;5,6}(7)\sim 2\centerdot 10^{-11}$, $\delta\varepsilon_{1;6,7}(7)\sim  1\centerdot 10^{-11}$,
$\delta\varepsilon_{2;1,2}(7)\sim 1\centerdot 10^{-11}$, $\delta\varepsilon_{2;2,3}(7)\sim 2\centerdot 10^{-11}$,
$\delta\varepsilon_{2;3,4}(7)\sim 9\centerdot 10^{-11}$,  $\delta\varepsilon_{2;4,5}(7)\sim 5\centerdot 10^{-11}$,
$\delta\varepsilon_{2;5,6}(7)\sim 5\centerdot 10^{-11}$, $\delta\varepsilon_{2;6,7}(7)\sim  3\centerdot 10^{-11}$.\\
A structure of the spectrum which includes two low energy HB has the following form for $\alpha=3$ as an example:\\
$\epsilon_{1,1}(3)=0.0017 \Longrightarrow \delta\varepsilon_{1;1,2}(3)\sim 5\centerdot 10^{-12}\Longrightarrow\epsilon_{1,2}(3)=0.0067 \Longrightarrow\delta\varepsilon_{1;2,3}(3)\sim 9\centerdot 10^{-11}\Longrightarrow
\epsilon_{1,3}(3)=0.0050 \Longrightarrow \Delta_{1,2}(3)= 0.1038 \Longrightarrow \epsilon_{2,1}(3)= 0.0066\Longrightarrow \delta\varepsilon_{2;1,2}(3)\sim 9\centerdot 10^{-13}\Longrightarrow \epsilon_{2,2}(3)= 0.0166\Longrightarrow \delta\varepsilon_{1;2,3}(3)\sim 9\centerdot 10^{-11}\Longrightarrow \epsilon_{2,3}(3)= 0.0100\Longrightarrow\Delta_{2,3}(3)=0.0820$.

Positive values  of $\delta\varepsilon_{j;i,i+1}(\alpha)$  correspond to quasigaps or zero density states at fraction fillings, since its values are extremely small $\sim 10^{-10}-10^{-13}$. The fine structure of the low energy HB does not change  at different $q$.
According to numerical calculations,  at a given $\alpha$ each HB is split by  quasigaps into $\alpha$ subbands, the fine structure of each HB is formed.
Thus, the spectrum is determined by two types of gaps, the gaps $\Delta_{j,j+1}$ which determine the insulator states of the system with an entire filling $\rho_j =\frac{j}{q}$ (where $j$ is integer) and $\delta\epsilon_{j;i,i+1}$ with a fractional filling in each HB $\nu=\frac{i}{\alpha}$ (here  $i=1,...,\alpha-1$), moreover $\Delta_{j,j+1}\gg \delta\epsilon_{j,i;j,i+1}$. Thus, the structure of the spectrum is preserved in a fairly wide range of values $\lambda$, as noted above. Most likely the quasigaps in the spectrum determine the points of tangency of the subbands, and therefore they are defined as quasigaps.

\section*{Fractional-filled steady states}

In this section we consider a stability of the fine structure of the Hofstadter spectrum. Let us fix the chemical potential which corresponds to the fractional filling of each HB (for $\alpha =2$, $\nu=\frac{1}{2}$  and $ \nu=1\frac{1}{2}$) and numerical calculate the energy of electron liquid for different rational fluxes, which corresponds integer $\alpha$ and $\alpha^{-1}$. A steady state corresponds to the minimum energy for given flling. The energy density as a function of $\alpha$ is shown in Fig 1 a). A steady state is realized at $\alpha=\frac{1}{20}$ for the first HB and $\alpha=\frac{1}{25}$ for the second, as result, the fine structure of the Hofstadter spectrum is unstable at $\alpha=2$ or $\frac{1}{2}$-fractional filling.
From numerical analyse of stability of fine structures at different $\alpha$ it follows that the fine structure of HB is stable when HB is  filled $\nu< \frac{1}{2}$. The point $\nu =\frac{1}{2}$ is similar to the point of the phase transition, therefore in this point  the behavior of the electron liquid is rather critical. In Figs 1 we presented also the calculations of energy density for $\nu=\frac{1}{4},\frac{3}{4}$ b),
$\nu=\frac{1}{3},\frac{2}{3}, 1\frac{1}{3}, 1\frac{2}{3}$ c), $\nu=\frac{1}{7},\frac{2}{7}, \frac{3}{7}$ d)  $\nu=1\frac{1}{7},1\frac{2}{7}, 1\frac{3}{7}$ e),  $\nu=2\frac{1}{7},2\frac{2}{7}, 2\frac{3}{7}$ f).
For steady fractional Hall states the minimum energy is reached at $\alpha_c=3$ for $\nu=(G-1)+\frac{1}{3}$, $\alpha_c=4$  for $\nu=(G-1)+\frac{1}{4}$, $\alpha_c=7$ for $\nu=(G-1)+\frac{1}{7}$,  $\nu=(G-1)+\frac{2}{7}$, $\nu=(G-1)+\frac{3}{7}$, here G is the HB number. For rational fluxes the energy density does not depend on the value of $\alpha$ at $\alpha>\alpha_c$. At $\alpha_c$ the Hubbard interaction shifts  HB (decreasing energy with compare homogeneous state) and increases the bandwidths of subbands (increasing the energy). The summary energy  occurs as a result of the competition of these terms, that are determined by $U$.

\begin{figure}[tp]
     \centering{\leavevmode}
\begin{minipage}[h]{.45\linewidth}
\center{
\includegraphics[width=\linewidth]{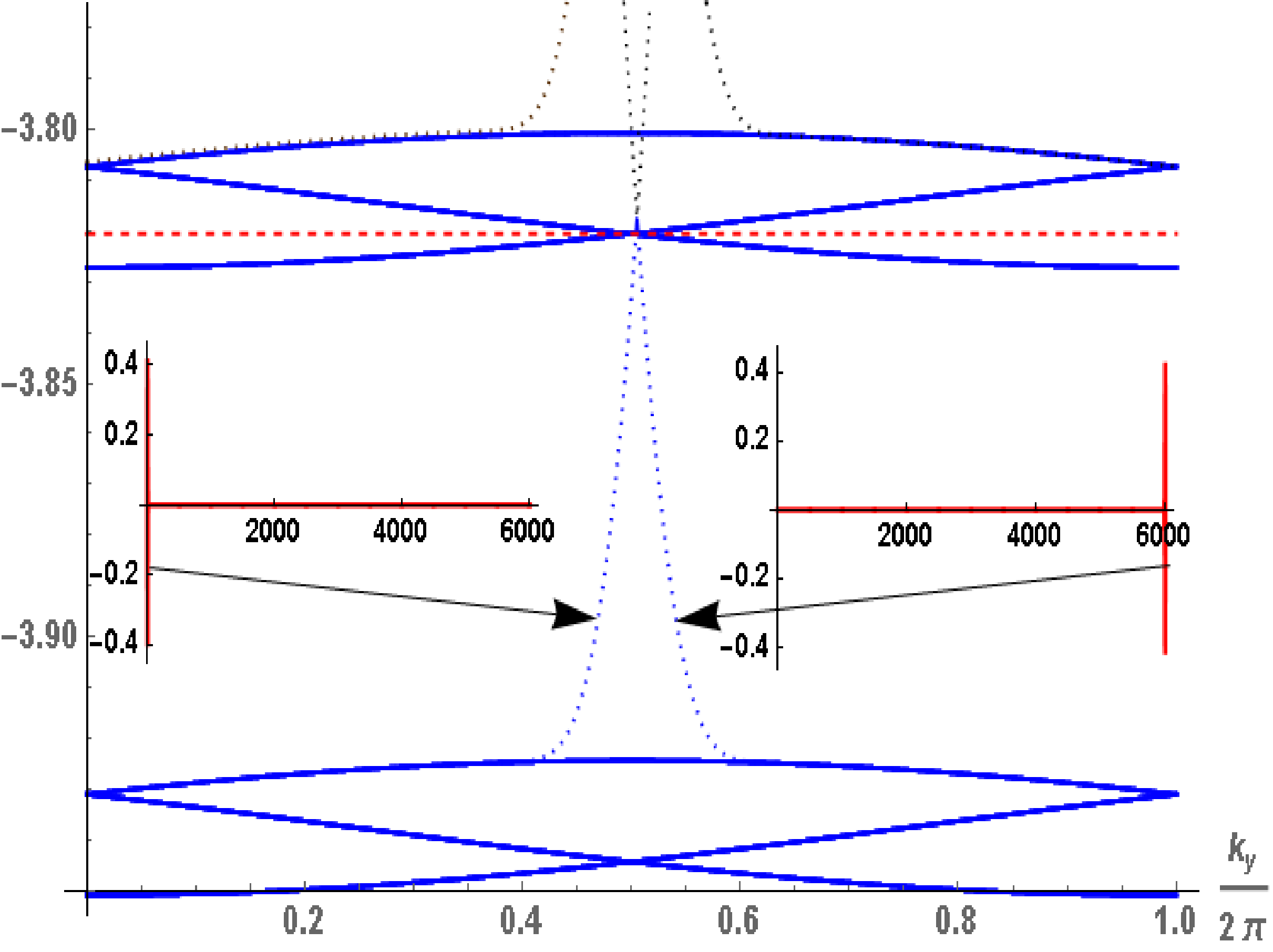} a)\\
                  }
    \end{minipage}
\begin{minipage}[h]{.45\linewidth}
\center{
\includegraphics[width=\linewidth]{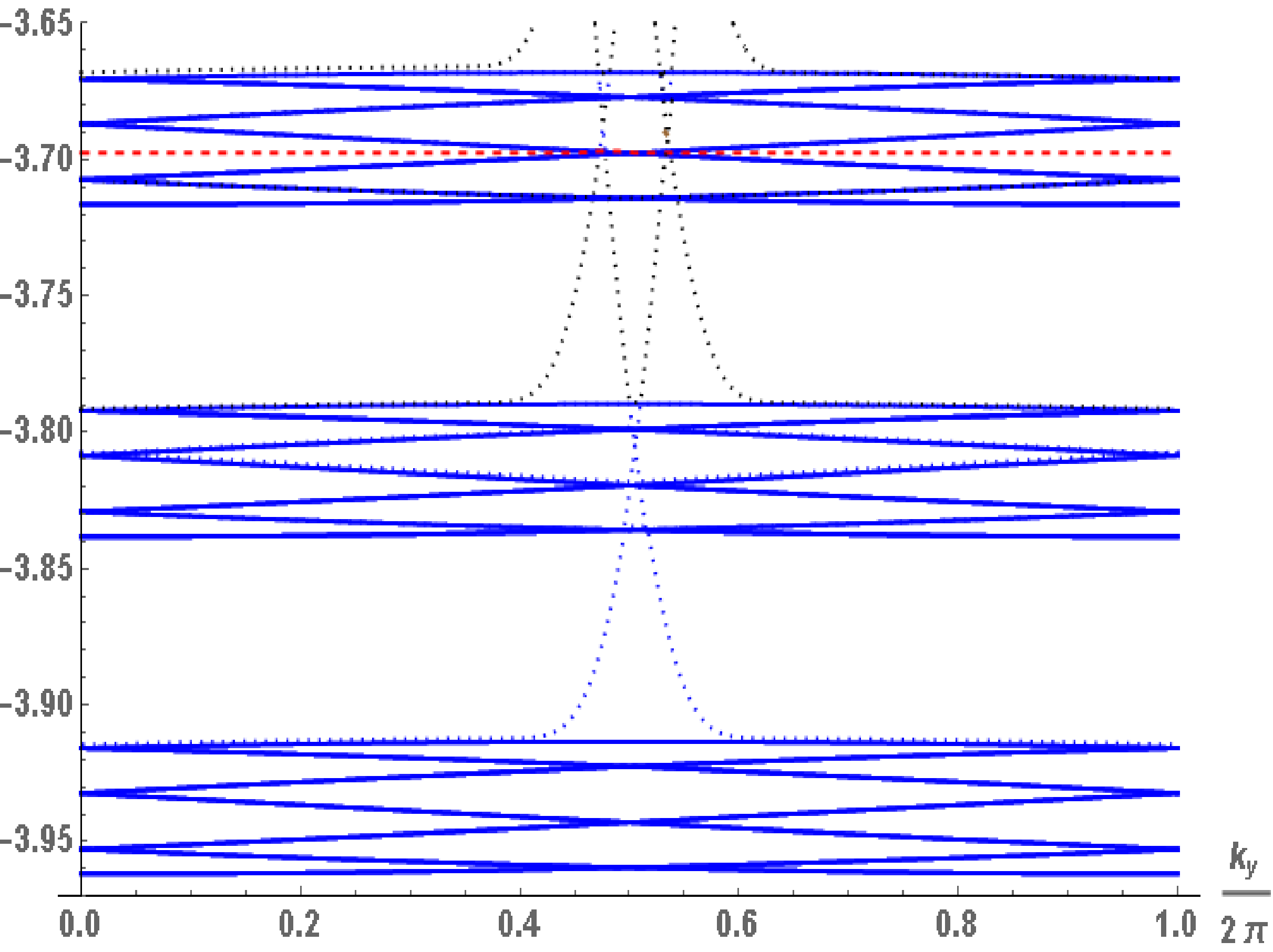} b)\\
                  }
    \end{minipage}
\caption{(Color online)
A fine structure of the two  a) and three b) lower energy HB calculated at $q = 100$, $U=1$, $\rho=1\frac{1}{3q}$ a) and $\rho=2\frac{3}{7q}$ b) for a sample in the form hollow cylinder with open boundary conditions along the y-direction, $k_y$ is the wave vector, red dashed lines denote the Fermi energies.
The dotted lines mark the dispersion of edge modes, the inserts illustrate them, where the amplitude of the wave function is calculated as a function of the x-coordinate  (1 and $6 \centerdot 10^3$ are the boundaries).
  }
\label{fig:2}
\end{figure}
\section*{Edge modes, fractional-Hall conductivity}

It is convenient to consider the behavior of edge modes when calculating quasiparticle excitations obtained for stripe geometry with open boundary conditions: the boundaries, parallel to the y-axis, are the edges of the hollow cylinder.
Let us focus on the calculation of two low energy HB at $\alpha=3$ and three HB at $\alpha=7$, we fix the Fermi energies which correspond fractional filling $\nu=1\frac{1}{3}$  and $ \nu =2\frac{3}{7}$. In the case $\alpha=3$ each HB splits on three subbands forming its fine structure. The quasigaps between there subbands extremely small (see in Fig.2 a)).  HB form edge modes in the forbidden region of the spectrum between them. These modes split from the upper and lower HB and are localized at the boundaries. Edge modes coexist with the fine structure of each HB, except for the first, in which they are not formed. In contrast to IQHE these modes connect also the nearest-neighbor subbands in the fine structure of each HB (except for the first). Extpemly small quasigaps do not kill topology of HB (the number of the edge modes is conserved at filling of HB).
The Hall conductance is determined by the same edge modes with different fraction filling $1+\frac{1}{3}$ for $\alpha =3$ and $2+\frac{3}{7}$ for $\alpha =7$.

\section*{Conclusions}

The Hofstadter model with short-range repulsion is considered within the mean-field approach, which allows one to study FQHE.
We described IQHE and FQHE using the same approach. and it is shown that \\
$\bullet$  short-range repulsion forms a steady fine structure of the Hofstadter spectrum when the filling of HB less than half;\\
$\bullet$  at partially filling of HB, subbands in the fine structure of HB are separated by extremely small quasigaps;\\
$\bullet$  these quasigaps  do not destroy the HB topology, only HB determines the number of the edge modes (the Chern number of HB is conserved);\\
$\bullet$  chiral edge modes located at the boundaries connect the nearest-neighbor subbands and determine the Hall conductivity with fractional filling;\\
$\bullet$  chiral edge modes are not formed into the first HB, therefore fractional conductance is not realized into the lowest (first) HB.

Numerical calculations were carried out in the semi-classical limit, which corresponds to the experimental conditions.

\section*{Author contributions statement}

I.K. is an author of the manuscript

\section*{Additional information}

The author declares no competing financial interests.

\end{document}